\documentclass[preprint,aps,pra]{revtex4-1}
\usepackage{amsmath}
\usepackage{amssymb}
\usepackage{graphicx}
\usepackage{color}
\usepackage[capitalize]{cleveref}

\newcommand{\ucd}[1]{{\stackrel{\triangledown}{#1}}}
\newcommand{\bs}[1]{\boldsymbol{#1}}
\newcommand{\ee}{\boldsymbol{e}}
\newcommand{\ud}[1]{\mathrm{d}#1}
\newcommand{\De}{\mathrm{De}}
\newcommand{\new}[1]{{#1}}
\newcommand{\remove}[1]{{}}
\newcommand{\edit}[1]{{#1}}

\begin{document}

\title{The mechanism of propulsion of a model microswimmer in a viscoelastic
    fluid next to a solid boundary}

\author{Thomas R. Ives}
\affiliation{SUPA, School of Physics and Astronomy, The University of Edinburgh,
    James Clerk Maxwell Building, Peter Guthrie Tait Road, Edinburgh, EH9 3FD,
    United Kingdom}

\author{Alexander Morozov}
\email{alexander.morozov@ph.ed.ac.uk}
\affiliation{SUPA, School of Physics and Astronomy, The University of Edinburgh,
    James Clerk Maxwell Building, Peter Guthrie Tait Road, Edinburgh, EH9 3FD,
    United Kingdom}

\begin{abstract}
In this paper we study swimming of a model organism, the so-called Taylor's
swimming sheet, in a viscoelastic fluid close to a solid boundary. This
situation comprises natural habitats of many swimming microorganisms, and while
previous investigations have considered the effects of both swimming next to a
boundary and swimming in a viscoelastic fluid, seldom have both effects been
considered simultaneously. We re-visit the small wave amplitude result obtained
by Elfring and Lauga (Gwynn J. Elfring and Eric Lauga, in Saverio E. Spagnolie,
editor, Complex Fluids in Biological Systems, Springer New York, New York, NY,
2015), and give a mechanistic explanation to the decoupling of the effects of
viscoelasticity, which tend to slow the sheet, and the presence of the boundary,
which tends to speed up the sheet.  We also develop a numerical spectral method
capable of finding the swimming speed of a waving sheet with an arbitrary
amplitude and waveform.  We use it to show that the decoupling mentioned above
does not hold at finite wave amplitudes and that for some parameters the
presence of a boundary can cause the viscoelastic effects to increase the
swimming speed of microorganisms.
\end{abstract}

\maketitle

\section{Introduction}

Many microorganisms are able to propel themselves through fluid environments by
deforming their bodies. The small size of these organisms, ranging from a few
micrometers in the case of most bacteria, to tens or hundreds of micrometers in
the case of eukaryotes, and their relatively small propulsion speeds dictate
that their swimming typically occurs in the low-Reynolds-number regime, and that
the fluid flow around them obeys the Stokes equation \cite{Lauga2009a}. As was
pointed out by Purcell \cite{Purcell1977b}, this poses a severe restriction on
how microorganisms move since they have to break the intrinsic
time-reversibility of the Stokes equation -- a result commonly known as
Purcell's scallop theorem. In order to propel, microorganisms have to either
deform their bodies or move parts of their bodies in a non-time-reversible
fashion, and a vast number of studies considered how various modes of propulsion
work and what the resulting properties of microorganisms' motion are (see
\cite{Brennen1977,Lauga2009a,Guasto2012} and references therein).

Arguably the most influential model of swimming at low Reynolds numbers is
Taylor's swimming sheet model \cite{Taylor1951a} that guided the later studies
of microorganism propulsion.  It comprises an infinite inextensible two
dimensional sheet, periodic in space, that can change its shape by propagating a
wave with a speed $c$ along its waveform. From the point of view of an external
observer, its shape traces out the curve $y_s$ in the $xy$-plane, given by
\begin{equation}
    y_s(x, t) = f(x-(c-U)t),
    \label{eq:waveform}
\end{equation}
where the wave is travelling in the positive $x$-direction, and the organism
swims at a speed $U$ along the same axis. The waveform $f(x)$ is a periodic
function with the period $2\pi/k$, where $k$ is the associated wavenumber. For
sheets with sinusoidal waveforms, $f(x) = b\sin(kx)$, and small amplitudes, $bk
\ll 1$, Taylor demonstrated that the sheet moves with the speed $U_{BN} =
c\,b^2k^2/2$ in the direction opposite to the direction of wave propagation
\cite{Taylor1951a}.

This result was extended by Katz \cite{Katz1974a} who considered a Taylor sheet
swimming next to a solid boundary and showed that to lowest order in $bk$ its
swimming speed is given by
\begin{equation}
    U_N = c\frac{b^2k^2}{2}\left(
        \frac{\sinh^2(hk)+h^2k^2}{\sinh^2(hk)-h^2k^2}
    \right),
    \label{eq:ukatz}
\end{equation}
where $h$ is the distance between the middle line of the organism and the
boundary. Notably, this speed is larger than $U_{BN}$, the swimming speed in the
bulk, for any finite value of $h$, although this conclusion relies on the
assumption that the organism keeps the same kinematics in the bulk and next to a
wall.  The works by Taylor \cite{Taylor1951a} and Katz \cite{Katz1974a} were
instrumental in guiding later studies of low-Reynolds propulsion of various
model swimmers, both in the bulk
\cite{Li2014,Alouges2008,Blake1971b,Juarez,Blake1971b,Drummond2006a,Gray1955a,Hancock1953a,Montenegro-Johnson2014a,Pak2009a,Pironneau2006,Purcell1997a,Reynolds1965a,Tuck1968a,Sauzade2011a,Sznitman2010,Tam2007,Wiezel2016,Taylor1952a,Elfring1994,Jalali2014a,Lighthill1952a}
and close to surfaces
\cite{Berke2008a,Katz1975,Lebois2012a,Lopez2014a,Montenegro-Johnson2014,Schulman2014,Trouilloud2008a,Crowdy2011}.

Another important extension of the Taylor's result was derived by Lauga
\cite{Lauga2007a} who studied a waving sheet swimming in the bulk of a
viscoelastic fluid. Lauga showed \cite{Lauga2007a} that for a range of
constitutive models, the small-amplitude swimming speed is given by
\begin{equation}
    U_B = c\frac{b^2k^2}{2}\left(\frac{1 + \beta\De^2}{1+\De^2}\right).
    \label{eq:ulauga}
\end{equation}
Here, $\beta = \eta_s/(\eta_s + \eta_p)$, with $\eta_s$ and $\eta_p$ being the
viscosity of the solvent and the polymer components, respectively, and $\De =
\lambda c k$ is the Deborah number of the problem, where $\lambda$ is the
longest relaxation time of the fluid. For the fixed kinematics of the organism,
this result suggests that viscoelasticity reduces the propulsion speed of a
small-amplitude sheet compared to its Newtonian value, reaching for large
Deborah numbers the limit \edit{$\beta U_{BN}$}. These conclusions
were extended to other swimmers
\cite{DeCorato2015,DeCorato2016,Espinosa-Garcia2013,Fu2007,Fu2008a,Fu2009a,Lauga2007b,Lauga2009,Lauga2014,Lauga2014a,Riley2017,Riley2015,Riley2014,Salazar2016,Teran2010a,Espinosa-Garcia2013,Thomases2014}
or fluids with different rheological properties
\cite{Velez-Cordero2013,Li2015,Du2012,Leshansky2009a,Man2015,Velez-Cordero2013,Montenegro-Johnson2013},
and were used as a motivation for experimental studies
\cite{Celli2009,Dasgupta2013a,Gagnon2014a,Gagnon2016,Ishijima1986a,Martinez2014,Keim2012a,Liu2011a,Shen2011a}.

While the previous studies provide understanding of how individual effects
influence microswimming (with a notable exception of
\cite{Yazdi2014,Yazdi2015,Li2014a,LiArdekani2017}), the actual ecological situation of many
microorganisms often comprises a combination of these effects. Examples range
from sperm moving in a mucus along the cervix wall \cite{Bansil1995,Gaffney2011a,Kirkman-Brown2011a} to
bacterial pathogens invading biofilms of different bacterial species
\cite{Houry2012}. The simplest model to study such systems should include both
viscoelasticity of the suspending fluid and the presence of a solid boundary,
i.e. be a combination of the effects discussed above. The first step in this
direction was taken by Elfring and Lauga \cite{Elfring2015} who calculated the
swimming speed of a small-amplitude Taylor sheet swimming next to a boundary in
an Oldroyd-B fluid. Surprisingly, the swimming speed they obtain,
\begin{equation}
    U = c\frac{b^2k^2}{2}\left(\frac{1 + \beta\De^2}{1+\De^2}\right)
    \left(\frac{\sinh^2(hk)+h^2k^2}{\sinh^2(hk)-h^2k^2}\right).
    \label{eq:small-speed}
\end{equation}
is simply a combination of the swimming speeds $U_{BN}$, $U_N$ and $U_B$,
i.e. the effects of viscoelasticity and
the boundary factorise. While \cref{eq:ukatz}, \cref{eq:ulauga}, and
\cref{eq:small-speed} are often cited, there is currently no simple
understanding of the corresponding effects.

The purpose of this work is to provide a mechanistic explanation of the
interplay between viscoelasticity of the fluid and the presence of a solid wall.
Our paper is organised as follows. In \cref{section:small-amplitude} we consider
a small-amplitude Taylor sheet model swimming next to a boundary in an Oldroyd-B
fluid. We re-derive the result obtained by Elfring and Lauga \cite{Elfring2015},
\cref{eq:small-speed}, and obtain explicit expressions for the velocity and
stress fields around the swimmer \new{which we will use to develop a
    small-amplitude physical mechanism in \cref{section:discussion}}.  In
\cref{section:large-amplitude} we develop a numerical method based on a spectral
representation of hydrodynamic fields to calculate the swimming speed of a
Taylor sheet of any amplitude and shape, and apply it to the situation discussed
above. Finally, in \cref{section:discussion}, we use the velocity and stress
fields calculated in the previous Sections to explain the origin of
\cref{eq:ukatz,eq:ulauga}, and \cref{eq:small-speed}, and discuss the emerging
mechanism of propulsion.

\section{Small-amplitude swimming: analytical solution}
\label{section:small-amplitude}

\begin{figure}
    \includegraphics[width=114.3mm]{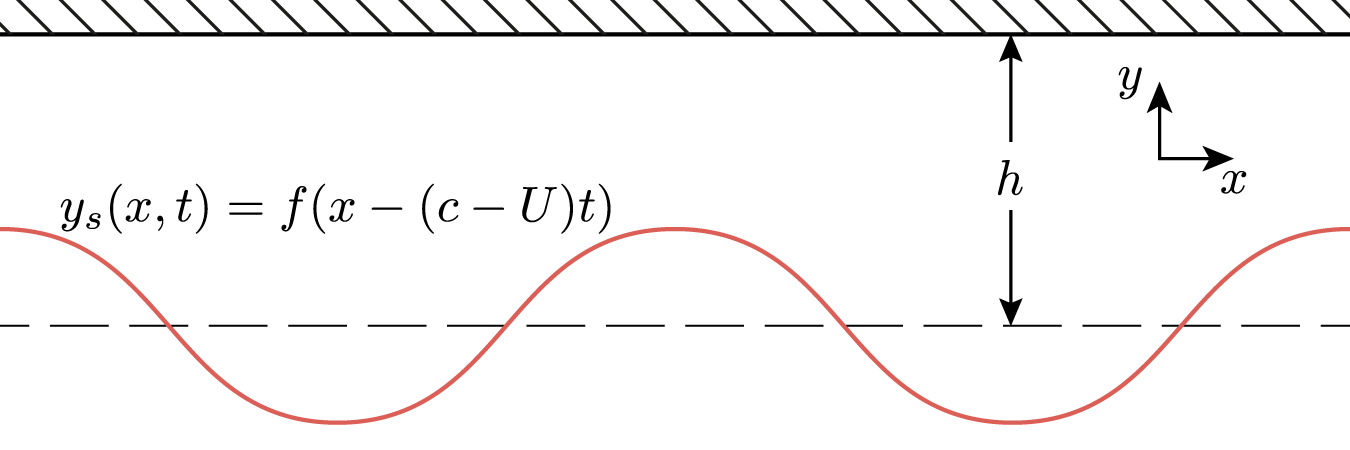}
    \caption{Schematic of a cross-section
        of a sheet a distance $h$ below a wall with the waveform $f(x)$.
    }\label{fig:sheet-wall}
\end{figure}

In this Section we consider a Taylor's waving sheet swimming in a viscoelastic
fluid next to a boundary in the small-amplitude limit, see
\cref{fig:sheet-wall}.  In the low-Reynolds-number limit, the flow of the fluid
around the organism is governed by the Stokes equation,
$\nabla\cdot\mathbf{\Sigma}=0$, where $\mathbf{\Sigma}$ is the total stress in
the fluid given by
\begin{equation}
    \mathbf{\Sigma} = -p\mathbf{1} + 2\eta_s\mathbf{D} + \bs{\tau}.
    \label{eq:total-stress}
\end{equation}
Here, $p$ is the pressure, $\mathbf{1}$ is the identity matrix, $\mathbf{D} =
(\nabla {\bf u} + \nabla {\bf u}^T)/2$ is the symmetric strain rate tensor, $\bf
u$ is the velocity of the fluid, $\eta_s$ is the solvent viscosity, and
$(\dots)^T$ denotes the transpose. The polymeric contribution to the total
stress, $\bs\tau$, arises due to the polymers being stretched and oriented by
local velocity gradients. Here we use one of the simplest viscoelastic
constitutive equations, the Oldroyd-B model \cite{bird1987dynamics,Morozov2015},
that develops large normal stresses in shear flows that are responsible for many
non-trivial effects exhibited by viscoelastic fluids \cite{bird1987dynamics};
but the model does not have any shear thinning effects, i.e. its material
properties are independent of local velocity gradients. Combined together, the
governing equations are given by
\begin{subequations}
    \label{seqs:governing}
    \begin{gather}
        -\nabla p + 2\eta_s \nabla\cdot\mathbf{D} +
        \nabla\cdot\bs{\tau} = \bs{0}, \label{eq:forcebal} \\
        \nabla\cdot\bs{u}  = 0, \label{eq:incomp} \\
        \bs{\tau} + \lambda \ucd{\bs{\tau}} = 2 \eta_p \mathbf{D},
        \label{eq:const}
    \end{gather}
\end{subequations}
where $\eta_p$ is the polymeric contribution to the fluid's viscosity, $\lambda$
is the longest relaxation time of the solution and we assumed the fluid to be
incompressible. The upper-convected Maxwell derivative is given by
\begin{align}
\ucd{\bs{\tau}} = \partial_t \bs{\tau} + \bs{u}\cdot\nabla
\bs{\tau} - \nabla\bs{u}^T\cdot\bs{\tau} - \bs{\tau}\cdot\nabla\bs{u}. \nonumber
\end{align}
The boundary conditions are given by the no-slip boundary conditions at the sheet
and the wall:
\begin{subequations}
    \label{seqs:bc-sheet-wall}
    \begin{gather}
        \left.\bs{u}\right|_{y=y_s} = \bs{u}_s, \\
        \left.\bs{u}\right|_{y=h} = \bs{u}_w,
    \end{gather}
\end{subequations}
where $\bs{u}_s$ and $\bs{u}_w$ are the velocity of the material points of the
sheet and the wall, respectively.

To address the situation sketched in \cref{fig:sheet-wall}, we solve a slightly
more general problem of the sheet in a channel with walls both above and below
it, placed at distances $h_+$ and $h_-$ from the centreline, respectively.  We
then return to the original single-wall problem by taking $h_+\to h$ and
$h_-\to\infty$.

We start by simultaneously introducing dimensionless variables and removing any
explicit time dependence with the help of the following transformation to
starred quantities
\begin{gather*}
    x^* = k(x - ct),\ y^* = k y,\ h_\pm^* = h_\pm k, \\
    U^* = \frac{U}{c},\ u^* = 1 + \frac{u}{c},\ v^* = \frac{v}{c}, \\
    p^* = \frac{p}{\eta c k},\ \bs{\tau}^* = \frac{\bs{\tau}}{\eta_p c k},
    \ \mathbf{\Sigma}^* = \frac{\mathbf{\Sigma}}{\eta c k}.
\end{gather*}
In these coordinates the velocity of the walls is $\bs{u}_w^*= -\ee_x$ and the
shape of the sheet is approximately fixed in time, such that $y^*_s(x^*) =
\epsilon \sin(x^*) + \mathcal{O}(\epsilon^3)$ where $\epsilon = bk$ is the
dimensionless wave amplitude.  Our goal then is to find the steady velocity
field surrounding the sheet and, from this, to calculate the sheet's swimming
speed.  From now on we will drop the $^*$s.

We expand the fields $p$, $\bs{u}$ and $\bs{\tau}$ into Taylor series about
$\epsilon = 0$.  For example, the pressure field is given by
\begin{equation}
    p = \sum_{n=0}^{\infty} p^{(n)} \epsilon^n.
\end{equation}
where $p^{(n)}$ is the ``$n$th-order'' contribution to the pressure field.

The velocity field is the solution to \cref{seqs:governing}, subject to the
no-slip boundary conditions of \cref{seqs:bc-sheet-wall}. This solution is
periodic in $x$, reflecting the symmetry of the underlying problem.  As we are
only interested in the lowest order of the small-$\epsilon$ expansion of the
swimming speed, we only need the boundary conditions to the lowest order in
$\epsilon$.  From Taylor's original paper \cite{Taylor1951a}, we can show that,
to the lowest order, the velocity of the material points of the sheet,
$\bs{u}_s$, are given in our coordinates\new{, where we are co-moving with the
    wave,} by
\begin{subequations}
\begin{gather}
    u_s = -1 + \mathcal{O}(\epsilon^2), \\
    v_s = -\epsilon \cos(x)+ \mathcal{O}(\epsilon^2).
\end{gather}
\end{subequations}

In our coordinate system, the swimming speed of the sheet can be found by
averaging the velocity field along the length of the sheet.  Thus, up to the
second order in $\epsilon$, the swimming speed of the sheet is given by
\begin{equation}
    U = \langle \left.u^{(1)}\right|_{y=y_s} \rangle \epsilon +
    \langle \left.u^{(2)}\right|_{y=0} \rangle \epsilon^2 +
    \mathcal{O}(\epsilon^3).
     \label{eq:speed-general}
\end{equation}
To the lowest order in $\epsilon$, averages over the line
$y=y_s=\epsilon\sin(x)$ are equal to averages over the line $y=0$, thus we take
the $\langle\rangle$ above as simple $x$-averages.

We find the first and second order velocity fields by substituting the Taylor
expansion of each of the fields into \cref{seqs:governing} and consider each
power of $\epsilon$ separately.  To the zeroth order, this procedure yields the
following set of equations
\begin{gather*}
    -\nabla p^{(0)} + \beta \nabla^2\bs{u}^{(0)} +
     (1 - \beta) \nabla\cdot\bs{\tau}^{(0)} = \bs{0}, \\
    \nabla\cdot\bs{u}^{(0)} = 0, \\
    \bs{\tau}^{(0)} = \mathbf{D}^{(0)}, \\
    \left.\bs{u}^{(0)}\right|_{y=y_s} =
    \left.\bs{u}^{(0)}\right|_{y=h_+} =
    \left.\bs{u}^{(0)}\right|_{y=-h_-} = -\ee_x,
\end{gather*}
which has the trivial solution $p^{(0)} = 0$, $\bs{\tau}^{(0)} = \bs{0}$, and
$\bs{u}^{(0)} = -\ee_x$.  Note that the zeroth order velocity field does not
contribute to the swimming speed of the sheet as the latter is given by the
difference between the average velocities of the fluid at the sheet and at the
wall, which vanishes at the zeroth order.

The first order velocity field is in fact the same for an Oldroyd-B fluid as for
a Newtonian one \cite{Lauga2007a}.  To demonstrate this, we consider the first
order equations:
\begin{subequations}
    \label{seqs:first}
    \begin{gather}
        -\nabla p^{(1)} + \nabla\cdot(2\beta \mathbf{D}^{(1)} +
        (1 - \beta) \bs{\tau}^{(1)}) = \bs{0},
        \label{eq:first-forcebal}\\
        \nabla\cdot\bs{u}^{(1)} = 0, \label{eq:first-incomp}\\
        (1 - \De~\partial_x)\bs{\tau}^{(1)} =
        \mathbf{D}^{(1)}, \label{eq:first-const}\\
        \left.\bs{u}^{(1)}\right|_{y=y_s} = -\cos(x)\ee_y, \
        \left.\bs{u}^{(1)}\right|_{y=h_+} =
        \left.\bs{u}^{(1)}\right|_{y=-h_-} = \bs{0}. \label{eq:first-bc}
    \end{gather}
\end{subequations}
Here we have used the previous solution, $u^{(0)} = -1$, in
\cref{eq:first-const}, and we have re-arranged \cref{eq:first-forcebal} using
$\nabla^2\bs{u}^{(1)} = 2\nabla\cdot\mathbf{D}^{(1)}$. Let $\mathcal{L}$ be the
linear operator defined by
\begin{equation}
    \mathcal{L}(\bs{a}) = (1-\De~\partial_x)\nabla\cdot\mathcal{E}\cdot\bs{a},
\end{equation}
where $\mathcal{E} = \ee_x\ee_y - \ee_y\ee_x$.  Applying $\mathcal{L}$ to
\cref{eq:first-forcebal}, we obtain
\begin{equation}
    \nabla\cdot\mathcal{E}\cdot\nabla\cdot\left(
        (1-\De~\partial_x)(2\beta \mathbf{D}^{(1)} +
        (1 - \beta) \bs{\tau}^{(1)})
    \right) =2\nabla\cdot\mathcal{E}\cdot\nabla\cdot\left(
        1 - \beta\De~\partial_x
    \right) \mathbf{D}^{(1)}= 0, \label{eq:first-curl}
\end{equation}
where we have used the commutativity of differential operators and
\cref{eq:first-const} to remove $\bs{\tau}^{(1)}$.  This equation is satisfied
either by a $\mathbf{D}^{(1)}$ for which
$\nabla\cdot\mathcal{E}\cdot\nabla\cdot\mathbf{D}^{(1)} = 0$, or by a
$\mathbf{D}^{(1)}$ for which  $(1 - \beta\De~\partial_x)\mathbf{D}^{(1)} = 0$.
The first of these conditions is satisfied by the Newtonian solution, while the
second has no non-trivial solutions which are periodic in $x$.  Moreover, since
the boundary conditions are the same as in the Newtonian case, we conclude that
the first order velocity field in an Oldroyd-B fluid is that same as its
Newtonian counterpart.

In his original analysis of a Taylor sheet swimming next to a wall, Katz
\cite{Katz1974a} showed that the first order velocity field, $\bs{u}^{(1)}_\pm$
in the Newtonian fluid above ($+$) and below ($-$) the sheet is given by
\begin{subequations}
    \label{seqs:first-order-vel}
    \begin{gather}
        u^{(1)}_\pm = (1 + A_\pm - B_\pm y)\sin(x)\sinh(y) +
        A_\pm y \sin(x)\cosh(y), \\
        v^{(1)}_\pm = -(A_\pm y + B_\pm)\cos(x)\sinh(y) -
        (1 - B_\pm y)\cos(x)\cosh(y),
     \end{gather}
\end{subequations}
where
\begin{gather}
        A_\pm = \frac{\sinh^2(h_\pm)}{\sinh^2(h_\pm) - h_\pm^2},~
        B_\pm = \frac{\pm \sinh(h_\pm)\cosh(h_\pm) \pm h_\pm}
        {\sinh^2(h_\pm) - h_\pm^2}.~ \nonumber
\end{gather}
The contribution of this field to the swimming speed is given by the first term of
\cref{eq:speed-general}, which reads
\begin{equation}
    \langle \left.u^{(1)}\right|_{y=y_s} \rangle \epsilon =
    \epsilon^2 (A + \frac{1}{2}) + \mathcal{O}(\epsilon^3).
\end{equation}
Here we have dropped the explicit $\pm$ notation, but we note that this
contribution to the swimming speed is different for each region of the fluid.
Below, we ensure that the total swimming speed is the same regardless of whether
we use the fluid velocity above or below the sheet; but first we calculate the
second term in \cref{eq:speed-general}.

The second order set of governing equations is given by
\begin{subequations}
    \label{seqs:second}
    \begin{gather}
        -\nabla p^{(2)} + \beta\nabla^2\bs{u}^{(2)} +
        (1 - \beta) \nabla\cdot\bs{\tau}^{(2)} = \bs{0},
        \label{eq:second-forcebal}\\
        \nabla\cdot\bs{u}^{(2)} = 0, \label{eq:second-incomp}\\
        (1 - \De~\partial_x)\bs{\tau}^{(2)} =
        \mathbf{D}^{(2)} - \De\Big[
        \bs{u}^{(1)}\cdot\nabla\bs{\tau}^{(1)} -
        (\nabla\bs{u}^{(1)})^T\cdot\bs{\tau}^{(1)} -
        \bs{\tau}^{(1)}\cdot\nabla\bs{u}^{(1)}
        \Big]. \label{eq:second-const}
    \end{gather}
\end{subequations}
We have left out the boundary conditions, which we will deal with later.  Note
that in the second term of \cref{eq:speed-general}, the $x$-average commutes with
the $y$-substitution, as the substitute is independent of $x$.  Thus, we only
have to solve the $x$-average of \cref{seqs:second} to find the second order
swimming speed.  Considering the $x$-averages of the $x$-component of
\cref{eq:second-forcebal} and the $xy$-component of \cref{eq:second-const}, we
have
\begin{subequations}
    \label{seqs:second-average}
    \begin{gather}
        \beta\partial_{yy} \langle u^{(2)}\rangle +
        (1-\beta)\partial_y\langle\tau_{xy}^{(2)}\rangle = 0, \\
    \begin{aligned}
        \langle\tau_{xy}^{(2)}\rangle =
        \partial_y\langle u^{(2)} \rangle -
    \De\big[
        &\langle u^{(1)}\partial_x\tau_{xy}^{(1)}\rangle +
        \langle v^{(1)}\partial_y\tau_{xy}^{(1)}\rangle - \phantom{a}\\
        &\langle D_{xy}^{(1)}(\tau_{xx}^{(1)}+\tau_{yy}^{(1)})\rangle +
        \langle \Omega_{xy}^{(1)}(\tau_{xx}^{(1)}-\tau_{yy}^{(1)})\rangle
    \big],
    \end{aligned}
    \end{gather}
\end{subequations}
where $\mathbf{\Omega} = (\nabla\bs{u}^T-\nabla\bs{u})/2$ is the vorticity
tensor, and we have ignored the $x$-averages of $x$-derivatives of $x$-periodic
functions, which must vanish.  Since the first order fields are known,
\cref{seqs:second-average} is simply an ordinary differential
equation for $\langle u^{(2)} \rangle$, the solution to which is given by
\begin{gather}
    \label{eq:second-sol}
    \langle u^{(2)}_\pm \rangle = E_\pm + F_\pm y +
    \frac{(1-\beta)\De^2}{4(1+\De^2)}\big[
        G_\pm\cosh(2y) + H_\pm\sinh(2y)
    \big],
\end{gather}
where
\begin{align}
& G_\pm = (2 + 4 A_\pm + A_\pm^2 - B_\pm^2) -
        4 B_\pm (1 + A_\pm) y + 2(A_\pm^2 + B_\pm^2) y^2, \nonumber \\
& H_\pm = 2 A_\pm (B_\pm + 2 (1 + A_\pm) y - 2 B_\pm y^2). \nonumber
\end{align}
Here, $E_\pm$ and $F_\pm$ are arbitrary constants.

%And thus the swimming speed of
%the sheet is given by
%\begin{equation}
%    \label{eq:second-speed-coeff}
%    U = \epsilon^2\left\{E_\pm + \frac{1 + \beta \De^2}{1 + \De^2}
%        (A_\pm + \frac{1}{2}) +
%    \frac{(1-\beta)\De^2}{4(1+\De^2)}(A_\pm^2-B_\pm^2)\right\},
%\end{equation}
%for either $+$ or $-$.

Until now, the solutions in the domains above and below the sheet were
completely independent. Their coupling is now ensured by applying appropriate
boundary conditions, which determine the constants $E_\pm$ and $F_\pm$. Similar
to the solution developed by Katz for swimming of a Taylor sheet next to a wall
in a Newtonian fluid \cite{Katz1974a}, we require that (i) the swimming speeds
we calculate from the upper and lower fluids match, and that (ii) the sheet is,
on average, force-free to second order in the $x$-direction.  The first order
flow field, which is the same as the first order Newtonian flow field, does not
apply a net-force to the sheet \cite{Katz1974a}, while the second order flow
field contributes a force
$\left.\mathbf{\Sigma}^{(2)}\cdot\bs{n}_\pm\right|_{y=0}$ where $\bs{n}_\pm =
\pm\ee_y + \mathcal{O}(\epsilon)$ is the inward-normal of the sheet in the
upper/lower fluid. To the second order, these boundary conditions are given by
\begin{subequations}
    \label{seqs:second-bc}
    \begin{gather}
        \left.\langle u^{(2)}_\pm \rangle\right|_{y=\pm h_\pm} = 0,\\
        \langle\left. u^{(1)}_+\right|_{y=y_s} \rangle\epsilon +
        \left.\langle u^{(2)}_+ \rangle\right|_{y=0}\epsilon^2 =
        \langle\left. u^{(1)}_-\right|_{y=y_s} \rangle\epsilon +
        \left.\langle u^{(2)}_- \rangle\right|_{y=0}\epsilon^2 +
        \mathcal{O}(\epsilon^3), \\
        \left.-\beta\partial_y\langle u^{(2)}_+\rangle\right|_{y=0}
        +\left.(1-\beta)\langle \tau^{(2)}_{xy,+}\rangle\right|_{y=0} =
        \left.-\beta\partial_y\langle u^{(2)}_-\rangle\right|_{y=0}
        +\left.(1-\beta)\langle \tau^{(2)}_{xy,-}\rangle\right|_{y=0},
    \end{gather}
\end{subequations}
which result in
\begin{subequations}
    \begin{gather}
        E_\pm = \frac{(1 + \beta\De^2)}{(1 + \De^2)}
        \frac{h_\pm(A_\mp - A_\pm)}{(h_+ + h_-)}
         + \frac{(1 - \beta)\De^2}{(1 + \De^2)}(A_\pm^2 - B_\pm^2), \\
        F_\pm = \frac{(1 + \beta\De^2)}{(1 + \De^2)}
        \frac{(A_\mp - A_\pm)}{(h_+ + h_-)}
         + \frac{(1 - \beta)\De^2}{h_\pm(1 + \De^2)}(
        J_\pm\cosh^2(h_\pm) + K_\pm\sinh^2(h_\pm)
        ),
\end{gather}
\end{subequations}
where
\begin{align}
& J_\pm = (1 - 2h_\pm B_\pm) + h_\pm^2 (A_\pm^2 - B_\pm^2), \nonumber \\
& K_\pm = (1 + 2A_\pm)(1 - 2h_\pm B_\pm) +
        (1 - h_\pm^2)(A_\pm^2 - B_\pm^2). \nonumber
\end{align}

Substituting the first and second order velocity fields into
\cref{eq:speed-general}, we finally arrive at
\begin{align}
    U &= \frac{\epsilon^2(1 + \beta \De^2)}{(h_+ + h_-)(1 + \De^2)}
    \Bigg[
    h_-\left(A_+ + \frac{1}{2}\right) + h_+\left(A_- + \frac{1}{2}\right)
    \Bigg] \nonumber \\
    \phantom{U} &= \frac{\epsilon^2(1 + \beta \De^2)}{2(h_+ + h_-)(1 + \De^2)}
    \Bigg[
    h_-\left(\frac{\sinh^2(h_+) + h^2_+}{\sinh^2(h_+) - h^2_+}\right) +
    h_+\left(\frac{\sinh^2(h_-) + h^2_-}{\sinh^2(h_-) - h^2_-}\right)
    \Bigg].
\end{align}
In the limit of $h_+\to hk$ and $h_-\to\infty$, we recover
\cref{eq:small-speed} as mentioned above.  The main implication of this result
is the observation that the effects of swimming next to a wall and swimming in
an Oldroyd-B fluid decouple at small wave amplitudes. The mechanism of this
decoupling is discussed in \cref{section:discussion}; but first we develop a
numerical method capable of calculating the swimming speed for any value of the
wave amplitude.

%%%%             %%%%       %%%%    %%%
%%%%           %%%  %%%     %%%%%   %%%
%%%%          %%%    %%%    %%% %%  %%%
%%%%         %%%%%%%%%%%%   %%%  %% %%%
%%%%%%%%%%   %%%      %%%   %%%   %%%%%
%%%%%%%%%%   %%%      %%%   %%%    %%%%

\section{Large-amplitude swimming: numerical method}
\label{section:large-amplitude}

Here we develop a numerical method to solve \cref{seqs:governing} subject to the
boundary conditions of \cref{seqs:bc-sheet-wall} for an arbitrary wave, with any
amplitude or shape.  As in the previous Section, we in fact solve the more
general problem of the sheet in a channel, with the walls above and below the
sheet a distance $h_+$ and $h_-$ from the centreline, respectively.  We perform
the following transformation to the starred variables
\begin{gather*}
    x^* = k(x - (c-U)t),\ y^* = k y,\ h_\pm^* = h_\pm k, \\
    U^* = \frac{U}{c},\ u^* = 1 - U^* + \frac{u}{c},\ v^* = \frac{v}{c}, \\
    p^* = \frac{p}{\eta c k},\ \bs{\tau}^* = \frac{\bs{\tau}}{\eta_p c k},
    \ \mathbf{\Sigma}^* = \frac{\mathbf{\Sigma}}{\eta c k},
\end{gather*}
that render the problem dimensionless. This transformation is different from the
small-amplitude  transformation because in this frame of reference the shape of
the sheet is exactly independent of time, as opposed to being independent of
time only in the limit of small wave amplitudes.  Again, we drop the $^*$s
in what follows.

To exploit the two-dimensional nature of the problem, we introduce a
stream-function $\psi(x, y)$, which is defined via its relationship to the flow
field $\bs{u}$: $u = \partial_y \psi$ and $v = -\partial_x \psi$.  This
substitution satisfies \cref{eq:incomp} for any choice of $\psi$. To
re-formulate \cref{seqs:governing} in terms of the stream-function, we take the
curl and divergence of  \cref{eq:forcebal} to obtain the complete set of
governing equations given by
\begin{subequations} % TODO: Check these
    \label{seqs:final-equations}
    \begin{gather}
        \beta \nabla^4 \psi - (1-\beta) \big[
            \partial_{xy}(\tau_{yy} - \tau_{xx}) +
            \Box^2 \tau_{xy}
        \big] = 0, \label{eq:bih} \\
        \nabla^2 p - (1-\beta) \big[
            \partial_{xx} \tau_{xx} +
            2\partial_{xy} \tau_{xy} +
            \partial_{yy} \tau_{yy}
        \big] = 0, \label{eq:lap} \\
        \tau_{xx} - 2\partial_{xy} \psi + \De\big[
            (\partial_y\psi\partial_x- \partial_x\psi\partial_y) \tau_{xx} -
            2\tau_{xx}\partial_{xy}\psi -
            2\tau_{xy}\partial_{yy}\psi
        \big] = 0, \label{eq:const-xx}\\
        \tau_{xy} + \Box^2 \psi  + \De\big[
            (\partial_y\psi\partial_x- \partial_x\psi\partial_y) \tau_{xy} +
            \tau_{xx}\partial_{xx}\psi -
            \tau_{yy}\partial_{yy}\psi
        \big] = 0,  \label{eq:const-xy}\\
        \tau_{yy} + 2\partial_{xy} \psi + \De\big[
            (\partial_y\psi\partial_x- \partial_x\psi\partial_y) \tau_{yy} +
            2\tau_{xy}\partial_{xx}\psi +
            2\tau_{yy}\partial_{xy}\psi
        \big]= 0, \label{eq:const-yy}
    \end{gather}
\end{subequations}
where $\Box^2 = \partial_{xx} - \partial_{yy}$, $\beta = \eta_s/(\eta_s +
\eta_p)$ is the viscosity ratio, and $\De = \lambda c k$ is the Deborah number of
the fluid. There are five differential equations, three of which are
non-linear, with five fields to solve for ($\psi$, $p$, $\tau_{xx}$,
$\tau_{xy}$, $\tau_{yy}$).

To solve \cref{seqs:final-equations}  numerically, we developed a spectral
method adapted for our geometry, where the two-dimensional stream-function,
pressure and polymeric stress fields are represented by Fourier-Chebyshev
series \cite{boyd2013chebyshev}. Since convergence properties of the
Fourier-Chebyshev basis are optimal in rectangular domains, we perform two
independent coordinate transformations, one for the fluid above the sheet and
the other for the fluid below, that project the corresponding domains onto
rectangular strips, periodic in one direction. These transformations from the
original coordinates $(x, y)$ to the new ones $(\eta_\pm,\xi_\pm)$ are given by
\begin{subequations}
    \label{seqs:transform}
    \begin{gather}
        \eta_\pm = x, \label{eq:transform-eta} \\
        \xi_\pm = 1 - 2\frac{\pm h_\pm - y}{\pm h_\pm - f(x)},
        \label{eq:transform-xi}
    \end{gather}
\end{subequations}
where ``$+$'' and ``$-$'' denote the regions above and below the sheet,
respectively.  In each domain, $\xi_\pm = 1$ corresponds to the domain's wall,
while $\xi_\pm = -1$ corresponds to the sheet, i.e.~the lower domain has been
flipped.  The two domains can be treated equivalently and from now on we will
drop the $\pm$ unless it is necessary. The solutions in these domains only
couple to each other through the boundary conditions at the sheet.

In each deformed domain $(\eta, \xi) \in [0,2\pi) \times [-1, 1]$, the
hydrodynamic variables are represented by truncated Fourier-Chebyshev series.
For example, the pressure field, $p$, is given by
\begin{equation}
    p(\eta, \xi) = \sum_{n=0}^{N-1}\sum_{m=0}^{M-1} p^{(nm)}
    F_n(\eta) T_m(\xi),
    \label{eq:pressure-field}
\end{equation}
where $T_m(\xi) = \cos(m\arccos(\xi))$ is the $m$th Chebyshev polynomial, and
$$
F_n(\eta) = \begin{cases}
            \sin(\frac{n + 1}{2} \eta) & n\;\text{odd}\\
            \cos(\frac{n}{2} \eta) & n\;\text{even},
            \end{cases}
$$ is the $n$th Fourier mode.  We choose the resolution $(N, M)$ such that the
error on truncating the series in \cref{eq:pressure-field} is small; for each
set of physical parameters this is assessed by increasing the resolution $(N,
M)$ until the value of the swimming speed of the sheet does not change by more
than $0.5\%$ between the two highest resolutions. Typically, this precision is
achieved by $N=33$ and $M=80$.

The spatial derivaties $\partial_\eta$ and $\partial_\xi$ of the
Fourier-Chebyshev representations are calculated by multiplying vectors
containing spectral coefficients with the $NM \times NM$ spectral derivative
matrices \cite{hussainibook,boyd2013chebyshev,Orszag1971}.  The spatial
derivatives in the original $(x, y)$-space are then trivially constructed with
the help of \cref{seqs:transform}, giving for each domain
\begin{subequations}
    \label{seqs:derivative}
    \begin{gather}
        \left(\frac{\partial}{\partial x}\right)_\pm =
        \frac{\partial}{\partial \eta_\pm} +
        (\xi_\pm - 1) \frac{f'(\eta_\pm)}{\pm h_\pm - f(\eta_\pm)}
        \frac{\partial}{\partial \xi_\pm},
        \label{eq:derivative-x}\\
        \left(\frac{\partial}{\partial y}\right)_\pm =
        \frac{2}{\pm h_\pm-f(\eta_\pm)}
        \frac{\partial}{\partial \xi_\pm}.
        \label{eq:derivative-y}
    \end{gather}
\end{subequations}
To calculate products of the fields represented in the spectral space, we use
the Fast Fourier transform \cite{hussainibook} with the
following collocation points
\begin{subequations}
    \label{seqs:points}
    \begin{gather}
        \eta_{n_c} = \frac{2\pi n_c }{N_c}, \label{eq:points-eta}\\
        \xi_{m_c} = \cos\left(\frac{\pi m_c}{M_c-1}\right), \label{eq:points-xi}
    \end{gather}
\end{subequations}
to evaluate the fields in the real space, calculate their product, and transform
the result back to the spectral space. Here, $n_c \in \left[0,N_c\right)$, $m_c
\in \left[0,M_c\right)$, and the collocation resolution $(N_c, M_c)$ is selected
to satisfy $N_c > 1.5 N$ and $M_c > 1.5 M$ in order to avoid aliasing issues
\cite{boyd2013chebyshev,hussainibook}.

Representing five governing equations in the truncated Fourier-Chebyshev basis
for each fluid domain yields a set of $10NM$ non-linear algebraic equations that
need to be complemented by the boundary conditions.  By using Fourier modes, we
have implicitly imposed periodic boundary conditions in the $\eta$-direction,
which correctly reflects the symmetry of the underlying problem.  We still need,
however, six boundary conditions (four for $\psi$ and two for $p$) along the
lines $\xi=\pm1$.  These boundary conditions are expanded in the Fourier basis
(as they are functions of $\eta$), generating $12N$ discretised boundary
conditions to substitute into the original set of $10NM$ discretised
governing equations.

The first boundary conditions to consider are the no-slip boundary conditions at
both the sheet and the wall, \cref{seqs:bc-sheet-wall}, where the velocities of
the material points of the sheet and the walls are given by \cite{Taylor1951a}
\begin{subequations}
    \label{seqs:mat-point-vel}
    \begin{gather}
        u_s(x) = - \frac{Q}{\sqrt{1 + f'(x)^2}},
        \label{eq:mpv-sheet-x}\\
        v_s(x) = - \frac{Qf'(x)}{\sqrt{1 + f'(x)^2}},
        \label{eq:mpv-sheet-y}\\
        \bs{u}_w = (U - 1)\ee_x,
        \label{eq:mpv-wall}
    \end{gather}
\end{subequations}
and
\begin{equation}
Q = \int_0^{2\pi}\sqrt{1 + f'(x)^2}\;\ud{x}. \nonumber
\end{equation}
The four boundary conditions are, therefore,
\begin{subequations}
    \label{seqs:bc-four}
    \begin{align}
        \left. \partial_y \psi \right|_{\xi = -1} &= u_s,
        \label{eq:vel-bc-sheet-x}\\
        \left. -\partial_x \psi \right|_{\xi = -1} &= v_s,
        \label{eq:vel-bc-sheet-y}\\
        \left. \partial_y \psi \right|_{\xi = 1} &= U - 1,
        \label{eq:vel-bc-wall-x}\\
        \left. -\partial_x \psi \right|_{\xi = 1} &= 0.
        \label{eq:vel-bc-wall-y}
    \end{align}
\end{subequations}
Note that the $x$-derivative of the $n=0$ Fourier mode vanishes, and that the
sheet's swimming speed $U$, which is unknown, appears in the $n=0$ mode of
\cref{seqs:bc-four}, thus different sets of boundary conditions are required for
the $n=0$ and the $n\ne0$ Fourier modes.  We address this below.  First, we
consider the other two boundary conditions required for the $n\ne0$ case.

As already mentioned above, we do not directly solve the force balance equation,
\cref{eq:forcebal}, but instead solve its derivatives (specifically its curl and
divergence, see \cref{seqs:final-equations}). The solutions to both problems may
differ, at most, by curl-free and divergence-free terms. To fix those terms, we
explicitly ensure that the force balance equation is satisfied at the
boundaries. Specifically, at both the sheet and the wall, we require that the
$\bs{n}\cdot\nabla\cdot\mathbf{\Sigma} = 0$, where $\bs{n}$ is the normal to the
surface. This yields the final two boundary conditions for the $n\ne0$ Fourier
modes:
\begin{subequations}
    \begin{multline}
        \big[f'(\eta) \partial_x p -\partial_y p +
            \beta  (f'(\eta)\partial_y - \partial_x) \nabla^2 \psi +
            \phantom{x}\\
            \phantom{x} + f'(\eta) \partial_x \tau_{xx} +
            (f'(\eta) \partial_y + \partial_x) \tau_{xy} +
            \partial_y \tau_{yy} \big]_{\xi = -1} = 0,
            \label{eq:forcebal-bc-sheet}
    \end{multline}
    \begin{equation}
        \big[ \partial_y p + \beta  \partial_x \nabla^2 \psi -
            \partial_x \tau_{xy} - \partial_y \tau_{yy} \big]_{\xi = 1} = 0,
            \label{eq:forcebal-bc-wall}
    \end{equation}
\end{subequations}
where $(0, -1)^T$ is the normal to the wall, and $(-f'(\eta), 1)^T$ the normal to
the surface of the sheet.

For the $n=0$ mode we replace
\cref{eq:vel-bc-sheet-x,eq:vel-bc-sheet-y,eq:vel-bc-wall-y,eq:forcebal-bc-wall}
with alternative boundary conditions.  First of all, we note that $\psi$ and $p$
are defined up to a constant as only their derivatives are physical, and we set
those constants to some arbitrary value. The other two boundary conditions
ensure that the average $x$-force being applied to each of the walls is zero.
And similarly to the small amplitude case, we have to couple the two domains by
requiring that the swimming speed of the sheet as calculated by each domain is
the same and that the sheet is a force-free swimmer.  Thus, we have
\begin{subequations}
    \label{seqs:zero-mode-bc}
    \begin{gather}
        \left. p\right|_{\xi=1} = 0, \label{eq:const-bc-p} \\
        \left.\psi\right|_{\xi=1} = 0, \label{eq:const-bc-psi}\\
        \big[ \beta \Box^2\psi - \tau_{xy}\big]_{\xi=1} = 0,
        \label{eq:force-bc-wall}\\
    \begin{split}
        \big[ f'(\eta_+) p_+ - 2\beta f'(\eta_+)\partial_{xy}\psi_+ -
        \beta\Box^2\psi_+ - f'(\eta_+)\tau_{xx, +} + \tau_{xy, +} \big]_{\xi_+=-1}
        = \\\big[ f'(\eta_-) p_- - 2\beta f'(\eta_-)\partial_{xy}\psi_- -
        \beta\Box^2\psi_- - f'(\eta_-)\tau_{xx, -} + \tau_{xy, -} \big]_{\xi_-=-1},
        \label{eq:force-bc-sheet}
    \end{split}\\
    \left.\partial_y\psi_+\right|_{\xi_+=1} =
    \left.\partial_y\psi_-\right|_{\xi_-=1}, \label{eq:vel-match-bc-wall}
    \end{gather}
\end{subequations}
where the absence of a $\pm$ implies that the boundary condition applies to
both domains.

In the spirit of the Chebyshev-tau method \cite{hussainibook}, for each Fourier
mode we replace the four highest Chebyshev modes of the discretised
\cref{eq:bih} and the two modes of \cref{eq:lap} with the boundary conditions
presented above. Combining everything together leads to the set of $10NM$
non-linear discretised equations, with the structure outlined in
\cref{tab:equations}.  With the solution to this set of equations,  the swimming
speed of the sheet is given by
\begin{equation}
    U = \left.\partial_y \psi \right|_{\xi=1} + 1.
\end{equation}

To actually solve this set of non-linear equations we use the Newton-Raphson
method \cite{hussainibook} with an analytically calculated Jacobian. In general,
for $\De > 0$, starting from an arbitrary initial guess does not lead to
convergence of the Newton-Raphson algorithm, and therefore, we employ a simple
continuation strategy. For each set of parameters, we start from the Newtonian
case, $\De = 0$, which is linear and can always be solved, and use its solution
as the initial guess for a slightly higher $\De$. This process is continued
until we either reach the target value of $\De$ or the algorithm fails to
converge, in which case a smaller step $\Delta\De$ is selected. In practice,
the $\Delta\De$ required for continuation becomes very small at sufficiently large
$\De$, leading to unreasonable computation times in which case we only report
the results up to that value of $\De$.

\begin{table}
    \centering
    \begin{tabular}{c||c|c}
                          & $n=0$ & $0 < n < N$ \\
        \hline
       \hline
        $0 \le m < M-4$   & \Crefrange{eq:bih}{eq:const-yy} &
                            \Crefrange{eq:bih}{eq:const-yy} \\
        \hline
        $m = M - 4$       & \begin{tabular}{@{}c@{}}
                                \Cref{eq:force-bc-sheet}(+)
                                \eqref{eq:vel-match-bc-wall}(--) \\
                                \Crefrange{eq:lap}{eq:const-yy}
                            \end{tabular} &
                            \begin{tabular}{@{}c@{}}
                                \Cref{eq:vel-bc-wall-y} \\
                                \Crefrange{eq:lap}{eq:const-yy}
                            \end{tabular} \\
        \hline
        $m = M - 3$       & \begin{tabular}{@{}c@{}}
                                \Cref{eq:force-bc-wall} \\
                                \Crefrange{eq:lap}{eq:const-yy}
                            \end{tabular} &
                            \begin{tabular}{@{}c@{}}
                                \Cref{eq:vel-bc-wall-x} \\
                                \Crefrange{eq:lap}{eq:const-yy}
                            \end{tabular} \\
        \hline
        $m = M - 2$       & \begin{tabular}{@{}c@{}}
                                \Cref{eq:const-bc-psi} \\
                                \Cref{eq:const-bc-p} \\
                                \Crefrange{eq:const-xx}{eq:const-yy}
                            \end{tabular} &
                            \begin{tabular}{@{}c@{}}
                                \Cref{eq:vel-bc-sheet-y} \\
                                \Cref{eq:forcebal-bc-wall} \\
                                \Crefrange{eq:const-xx}{eq:const-yy}
                            \end{tabular} \\
        \hline
        $m = M - 1$       & \begin{tabular}{@{}c@{}}
                                \Cref{eq:vel-bc-sheet-x} \\
                                \Cref{eq:forcebal-bc-sheet} \\
                                \Crefrange{eq:const-xx}{eq:const-yy}
                            \end{tabular} &
                            \begin{tabular}{@{}c@{}}
                                \Cref{eq:vel-bc-sheet-x} \\
                                \Cref{eq:forcebal-bc-sheet} \\
                                \Crefrange{eq:const-xx}{eq:const-yy}
                            \end{tabular}
    \end{tabular}
    \caption{Outline of how the $10NM$ descretised equations are constructed
        from the differential equations in \cref{seqs:final-equations} and the
        various boundary conditions \crefrange{seqs:bc-four}{seqs:zero-mode-bc}.
        \label{tab:equations}}
\end{table}

\begin{figure}
\begin{centering}
\includegraphics[width=3.2in]{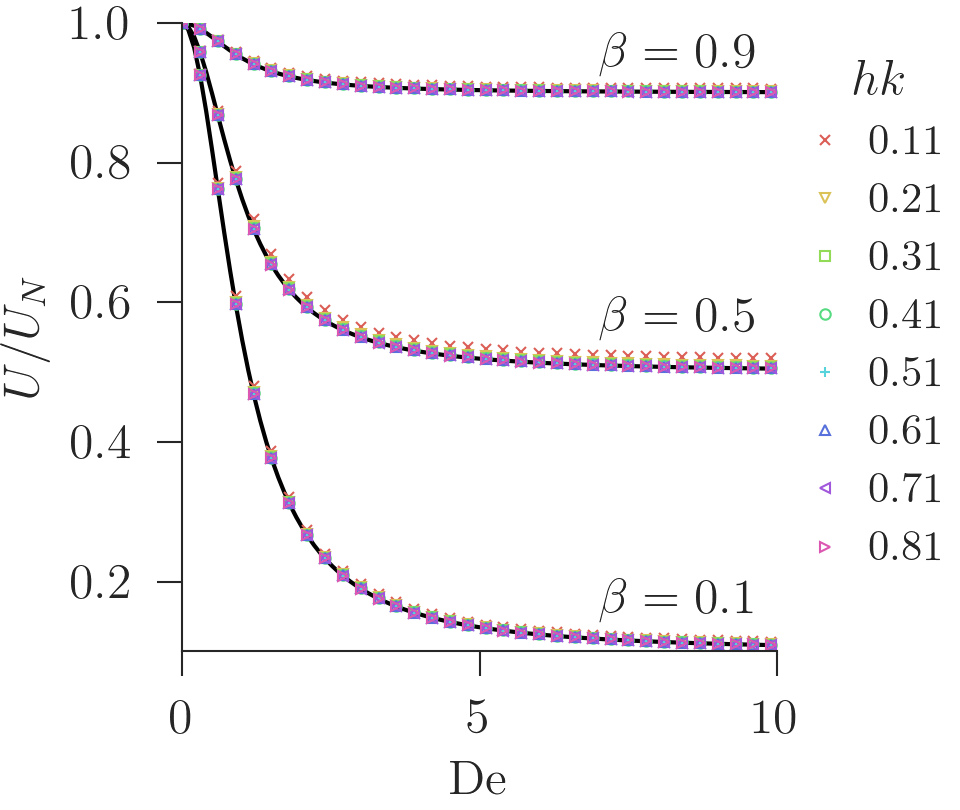}
\end{centering}
\caption[]{The swimming speed $U$ of a small-amplitude ($bk = 0.01$) Taylor
    sheet swimming next to a wall as the function of the Deborah number $\De$
    for various values of the solvent viscosity ratio $\beta$ and the distance
    from the wall $h$. The swimming speeds are normalised by the swimming speed,
    $U_N$, of the same geometric situation in a Newtonian fluid of the same
    viscosity.  The symbols are the results of our numerical calculations, while
    the solid black lines are the small amplitude predictions from
    \cref{eq:small-speed}.}
\label{fig:vsmallb}
\end{figure}

We verify that our numerical method correctly reproduces the
small-amplitude prediction \cref{eq:small-speed}. In \cref{fig:vsmallb} we plot
the swimming speed for a sheet with $bk = 0.01$ as a function of the Deborah
number $\De$ for various distances to the wall and viscosity ratios. As
expected, for this amplitude the numerically computed swimming speeds (symbols)
agree well with the analytical prediction of  \cref{eq:small-speed} (solid
line), again demonstrating that the effects of swimming next to a wall decouple
from the effects of swimming in a viscoelastic fluid at small amplitudes.

\begin{figure}
\centering \includegraphics[width=6.4in]{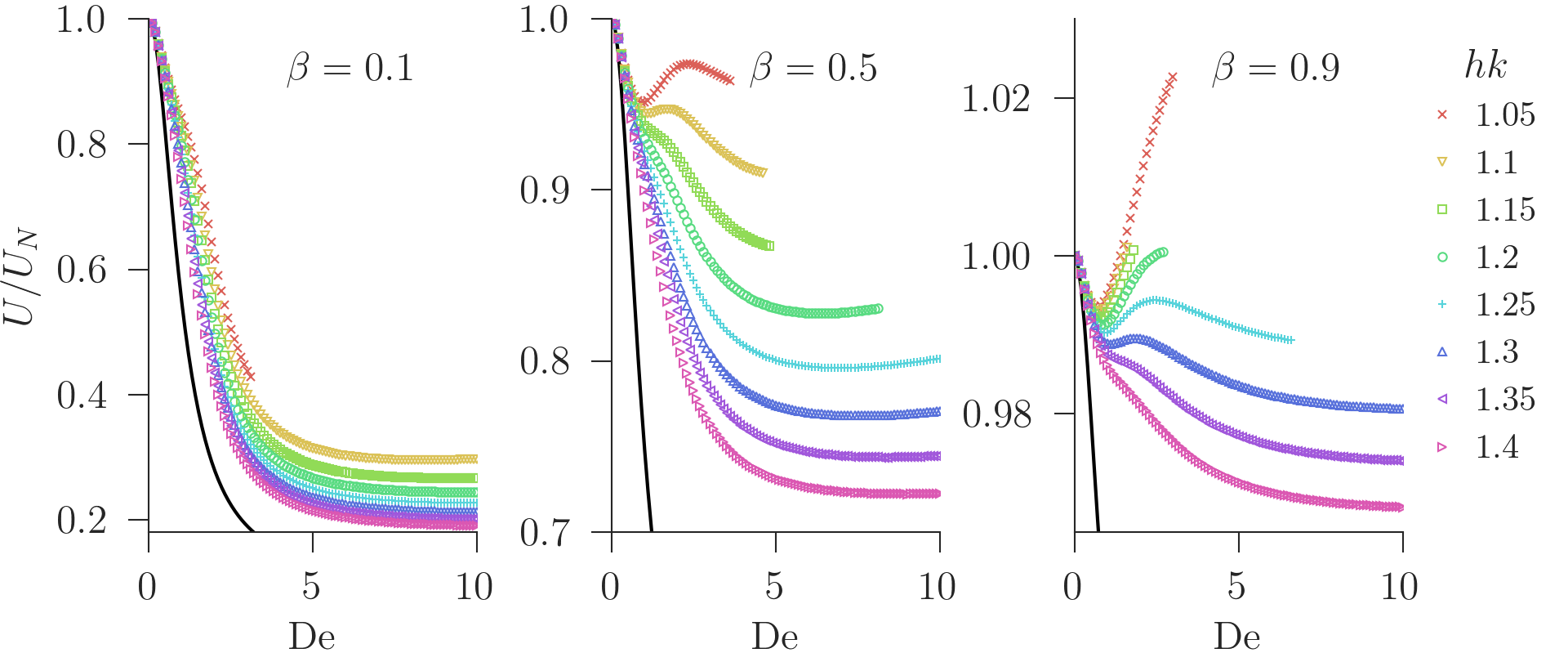}
% \begin{minipage}{0.49\textwidth}
% \includegraphics[width=\textwidth]{b0_5B0_1.png} \end{minipage}
% \begin{minipage}{0.49\textwidth}
% \includegraphics[width=\textwidth]{b0_5B0_5.png} \end{minipage}
% \begin{minipage}{0.49\textwidth}
% \includegraphics[width=\textwidth]{b0_5B0_9.png} \end{minipage}
\caption{The swimming speed $U$ of a finite-amplitude ($bk = 0.5$) Taylor sheet
    swimming next to a wall as the function of the Deborah number $\De$ for
    various values of the solvent viscosity ratio $\beta$ and the distance from
    the wall $h$. The swimming speeds are normalised by the swimming speed,
    $U_N$, of the same geometric situation in a Newtonian fluid of the same
    viscosity.  The symbols are the results of our numerical calculations, while
    the solid black lines are the small amplitude predictions from
    \cref{eq:small-speed}.}
\label{fig:large-b}
\end{figure}

The situation changes significantly for finite values of the wave amplitude. In
\cref{fig:large-b} we plot the swimming speed for a sheet with $bk=0.5$ as a
function of the Deborah number $\De$ for various distances to the wall and
viscosity ratios. We observe that the numerical data now deviates significantly
from the small-amplitude prediction \cref{eq:small-speed}. Despite this
deviation, for most values of $h$ and $\beta$ the swimming speed follows the
same trend as predicted by \cref{eq:small-speed}: starting from its Newtonian
value, it decreases monotonically with $\De$ and reaches a plateau value at
large Deborah numbers.
\new{However, for sufficiently small $h$ ($hk =1.05, 1.1$) at $\beta = 0.5$ and
    $\beta = 0.9$ the swimming speed breaks this trend and exhibits a
    non-monotonic dependence on $\De$.  This effect seems to be the stronger for
    larger values of $\beta$, which corresponds to more dilute solutions,
    reaching speeds faster than the Newtonian case for $\beta = 0.9$.  Also,
    there are indications that at lower values of $\beta$ the swimming speed
    starts to increase with $\De$ at sufficiently large values of the Deborah
    number.}
These results are further discussed in the
next Section.

\section{Discussion}
\label{section:discussion}

As we have demonstrated above, at small wave amplitudes the influence of the
polymeric stress on the swimming speed is the same for both swimming in the bulk
and next to a wall. In other words, the effects of the boundary and polymers
decouple and the swimming speed is given by the product of the corresponding
contributions, see \cref{eq:small-speed}. Let us discuss the mechanism of this
behaviour.

\begin{figure}
    \centering
    \includegraphics[width=114.3mm]{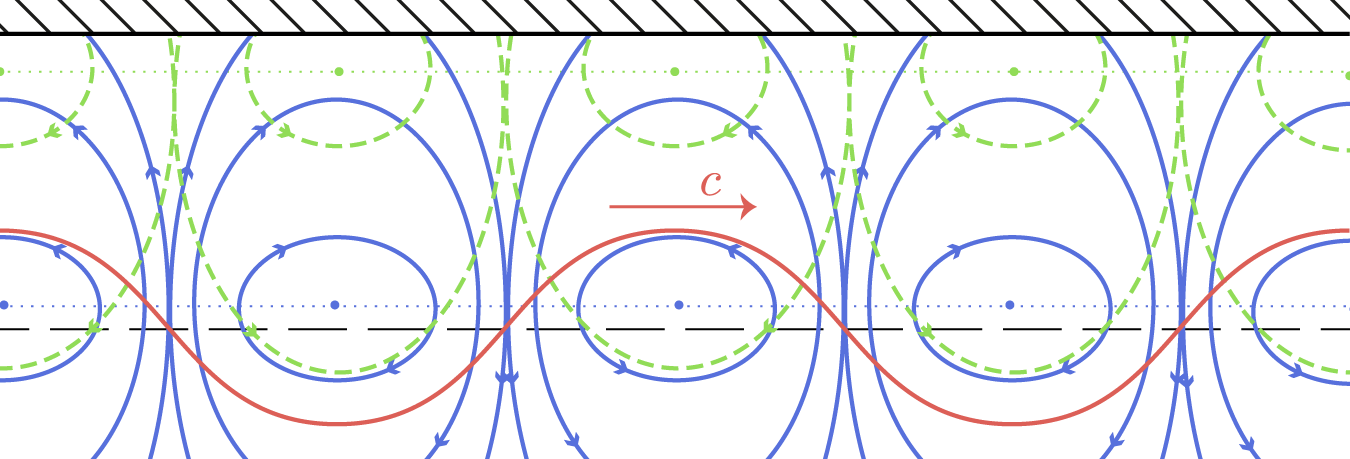}
    \caption{Gedankenexperiment demonstrating the sheet vortices (blue isolines)
        generated by the small-amplitude vertical motion of the material points
        of the sheet, and the velocity field it generates at an imaginary
        surface, \new{as the wave travels to the right}.  Note that this
        velocity field does not satisfy the no-slip boundary conditions. The
        surface velocity is cancelled by the wall vortices (green isolines), as
        discussed in the text.\label{fig:vortices}}
\end{figure}

We start by considering the kinematics of a Taylor sheet swimming in the bulk of
a Newtonian fluid. As noted by Taylor \cite{Taylor1951a} and Lauga and Powers
\cite{Lauga2009a}, at small wave amplitudes every point of the sheet is
oscillating approximately up and down, generating locally a vertical motion of
the surrounding fluid. Along one period of the sheet's waveform, for every point
moving up there is another point moving down with the same speed. Since the
fluid is incompressible, this sets an array of counter-rotating vortices along
the sheet, two vortices per period, see \cref{fig:vortices}. We will be
referring to them as the `sheet vortices'. As can be seen from
\cref{fig:vortices}, the presence of these vortices implies a velocity component
along the surface of the sheet, which on average drags the sheet along the
horizontal direction.

When this configuration of sheet vortices is placed next to a solid boundary, as
shown in \cref{fig:vortices}, it generates a non-zero velocity at the boundary
which obviously does not satisfy the no-slip boundary condition at this
boundary. This velocity is cancelled by the creation of an array of vortices
localised at the boundary with the same periodicity as the sheet vortices.
Along the boundary, the velocity of these `wall vortices' has the same magnitude
but opposing direction of the velocity of the sheet vortices. And thus, the
no-slip boundary condition is satisfied for the total velocity field. In turn,
the wall vortices have a contribution along the surface of the sheet that
effectively increases the speed of the sheet vortices, which in turn causes the
sheet to speed up. \new{Effectively, this implies that a small-amplitude sheet swimming
next to a wall can be viewed as a free-swimming sheet with faster sheet vortices.}

This argument is further corroborated by rearranging the first order velocity
field \cref{seqs:first-order-vel} in the following form, with dimensional
quantities
 \begin{gather*}
    \begin{split}
    u^{(1)} = \frac{c}{2}((A+B)ky-A-1)\sin(k(x-ct))\exp(-ky) +\\
              \frac{c}{2}((A-B)ky+A+1)\sin(k(x-ct))\exp(ky),
    \end{split}\\
    \begin{split}
    v^{(1)} = \frac{c}{2}((A+B)ky+B-1)\cos(k(x-ct))\exp(-ky) +\\
              \frac{c}{2}((B-A)ky-B-1)\cos(k(x-ct))\exp(ky),
    \end{split}
\end{gather*}
where we have dropped the $\pm$, as the fluid domains are equivalent and the
distinction between them is unimportant.  In the upper domain, the terms
proportional to $\exp(-ky)$ and $\exp(ky)$ correspond to the sets of vortices
which are localised at the sheet and at the wall, respectively.  The centres of
these vortices are located along the lines $ky=\omega_s$ and $ky=\omega_w$,
where
\begin{gather*}
    \omega_s = \frac{A+1}{A+B} =
    \frac{2\sinh^2(hk)-h^2k^2}{hk+\sinh(hk)\cosh(hk)+\sinh^2(hk)}, \\
    \omega_w = \frac{A+1}{B-A} =
    \frac{2\sinh^2(hk)-h^2k^2}{hk+\sinh(hk)\cosh(hk)-\sinh^2(hk)}.
\end{gather*}
For $hk>1$, $\omega_s\ll hk$ and $\omega_w\approx hk$, thus justifying sheet and
wall label of the arrays of vortices. Note that the vertical location
of the sheet vortices is moved from its bulk location, $ky=0$, to the line
$ky=\omega_s\ll hk$.  \remove{At small wave amplitudes, the effect of this displacement
is minor, while we suspect it plays an important role in finite-amplitude
swimming, see below.}

Now we turn to the effect that the viscoelasticity of the fluid has on the
swimming speed. We have recently studied the mechanism of the polymer-induced
slowing down of a Taylor sheet in the bulk \cite{Ives2017}, that we briefly
summarise here. As we have seen in \cref{section:small-amplitude}, the
first order velocity field is the same for a Newtonian and Oldroyd-B fluids, and
we start by considering its effect on the polymeric stresses. While this
vortical velocity field is locally a simple shear flow in most parts of the
domain, the regions in-between the vortices correspond to locally extensional
flows around stagnation points, similar to the ones observed in PIV measurements
by Shen and Arratia \cite{Shen2011a}. These extensional flows generate large
normal components of the polymer stresses, $\tau_{xx}$ and $\tau_{yy}$, that are
advected downstream by the fluid flow and, in turn,  generate significant shear
stresses in the fluid that drive the sheet horizontally, in the direction of the
wave propagation \cite{Ives2017}. \remove{Since the Taylor sheet swims in
the direction opposite to the direction of its wave propagation, this results in
slowing down of the sheet compared to a Newtonian fluid. We note that for
swimmers that travel in the direction of the wave propagation, this effect would
enhance their propulsion speed, see also \cite{Riley2015}.} \new{See also \cite{Riley2015} for a  discussion of the interaction between the multiple waves propagating in the opposite directions and viscoelasticity.}

Next to a boundary, the same mechanism applies at small wave amplitudes, since,
as we have argued above, a Newtonian sheet swimming next to a wall is equivalent
to a free-swimming sheet with faster sheet vortices and, hence, with a larger
swimming speed given by \cref{eq:ukatz}. This `effective' free-swimming sheet
would experience the same slowing down as discussed above and would swim with
the speed set by \cref{eq:ulauga}, where the Newtonian swimming speed $c b^2
k^2/2$ should be replaced with \cref{eq:ukatz}, arriving finally at
\cref{eq:small-speed}. \new{This is the fundamental reason behind the factorisation of 
the effects of viscoelasticity and the boundary.}

Now we turn to the case of finite-amplitude swimming. As shown in
\cref{fig:large-b}, at $bk=0.5$ the swimming speed deviates significantly from
the small amplitude prediction of \cref{eq:small-speed}.  Although, in the
majority of cases the trend of the velocity decreasing with the Deborah number
persists, the high-$\De$ value of the swimming speed $U_\infty$ is larger than
the asymptotic prediction of \cref{eq:small-speed}, \edit{$U_\infty > \beta U_N$},
and increases as the boundary is brought closer to the swimmer.  Moreover, in
some cases the swimming speed no longer decreases monotonically and can even
increase to swimming speeds greater than the Newtonian value.

To understand this behaviour we analyse the spatial distribution of the elastic
stresses in the fluid. Additionally, we plot the flow type parameter $\chi$,
defined as $\chi =
\frac{|\mathbf{D}|-|\mathbf{\Omega}|}{|\mathbf{D}|+|\mathbf{\Omega}|}$
\cite{Muller2007}.  Based on the invariants of the velocity gradient tensor, it
is designed to determine velocity type at every point in space, independent of
its local orientation: $\chi=1$ corresponds to purely extensional,  $\chi=0$ --
to shear, and $\chi=-1$ -- to purely rotation flows. Note that the flow type
parameter does not measure the magnitude of the flow, only its topology.

\begin{figure}
    \centering
    \begin{minipage}{0.33\textwidth}
    \includegraphics[width=2.1in]{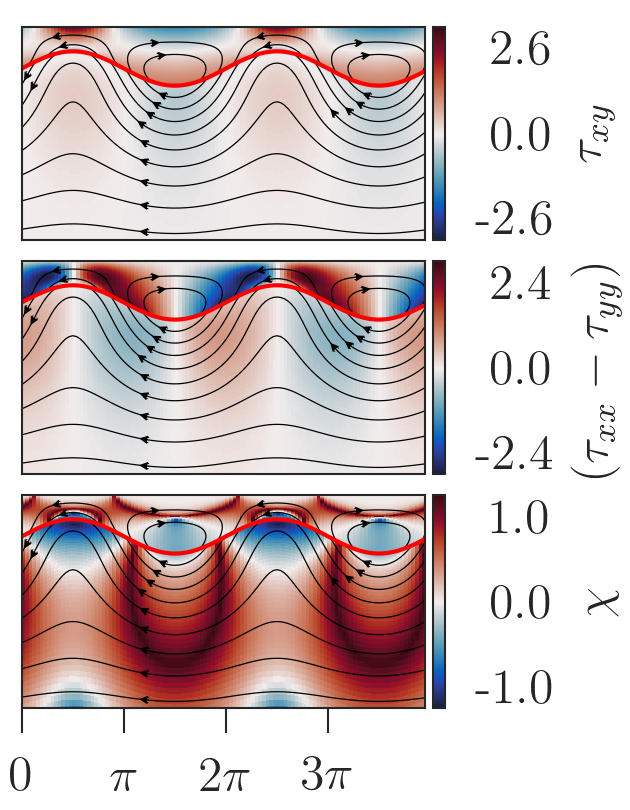}
    \end{minipage}
    \begin{minipage}{0.33\textwidth}
    \includegraphics[width=2.1in]{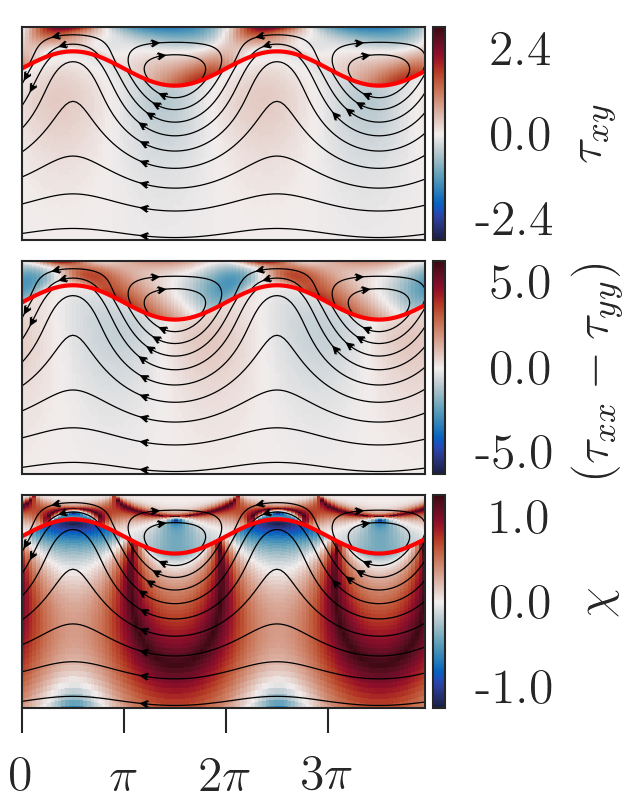}
    \end{minipage}
    \begin{minipage}{0.32\textwidth}
    \includegraphics[width=2.1in]{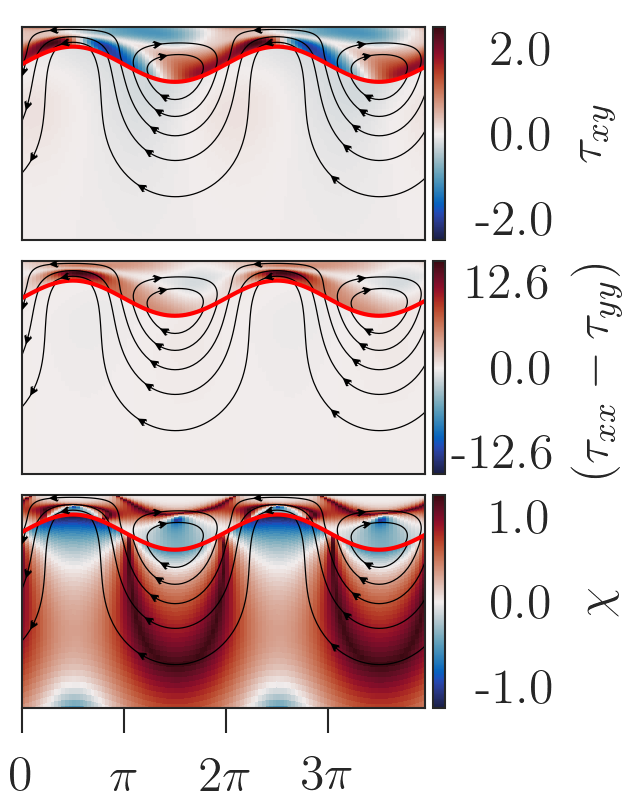}
    \end{minipage}
    \caption{The polymeric stress, $\bs{\tau}$ surrounding a sheet with
        amplitude $bk = 0.5$ near a wall a distance $h_+k = 1.2$ above it in an
        Oldroyd-B fluid with $\beta = 0.5$ and $\De = 0.0$ (left), $\De = 0.5$
        (middle) and $\De = 3.2$ (right).  The wall below the sheet is at $h_-k
        = 13.0$ which is far enough away to have no effect, however, the fluid
        domain is only shown until $ky = -5.0$.  The swimming speed of the
        sheet in each situation is $U = U_N = 0.362c$ (left), $U = 0.959U_N$
        (middle), $U = 0.854U_N$ (right).} \label{fig:tauchi}
\end{figure}

First we consider the case of moderate viscosity ratio and distance to the wall,
$\beta=0.5$ and $h_{+} k = 1.2$. In  \cref{fig:tauchi} we plot the shear stress
$\tau_{xy}$, the difference between normal stresses $\tau_{xx}-\tau_{yy}$, and
the flow type parameter for the Newtonian case $\De=0$ and two values of the
Deborah number, $\De=0.5$ and $3.2$, corresponding to the monotonic decrease of
the swimming speed from its Newtonian value. The plots are superimposed with the
isolines of the stream-function $\psi$ (black lines) and the local direction of
the velocity field (black arrows). Note that we plot the total velocity field
but will discuss it as consisting of the sheet and wall vortices, when useful.

\begin{figure}
    \centering
    \begin{minipage}{0.33\textwidth}
    \includegraphics[width=2.1in]{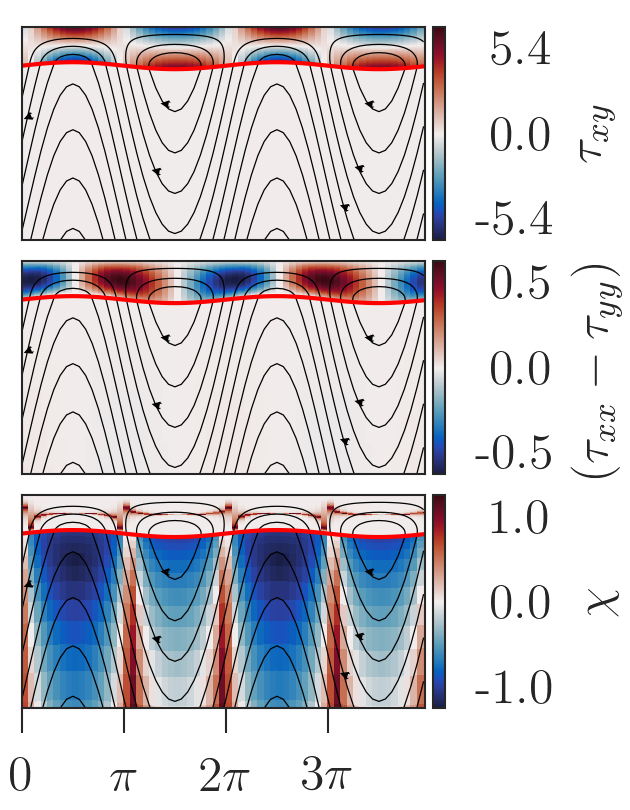}
    \end{minipage}
    \begin{minipage}{0.33\textwidth}
    \includegraphics[width=2.1in]{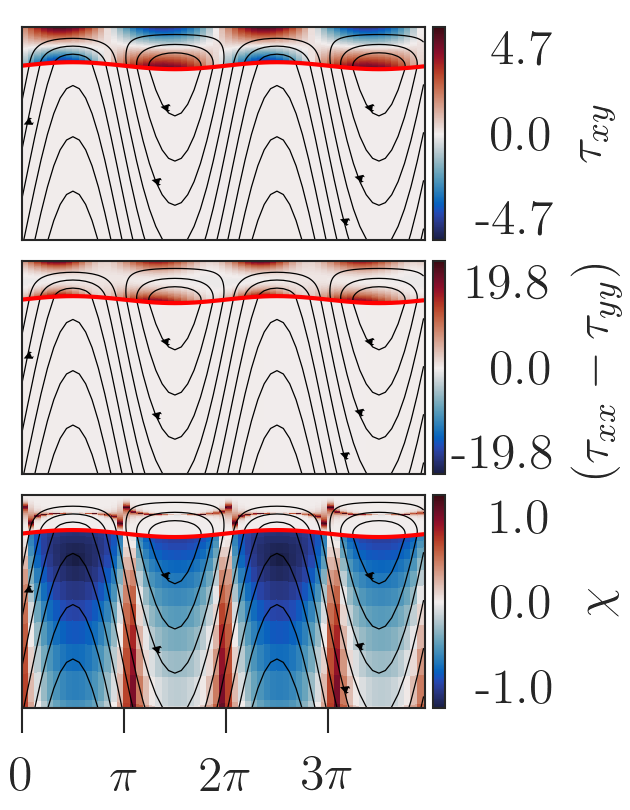}
    \end{minipage}
    \begin{minipage}{0.32\textwidth}
    \includegraphics[width=2.1in]{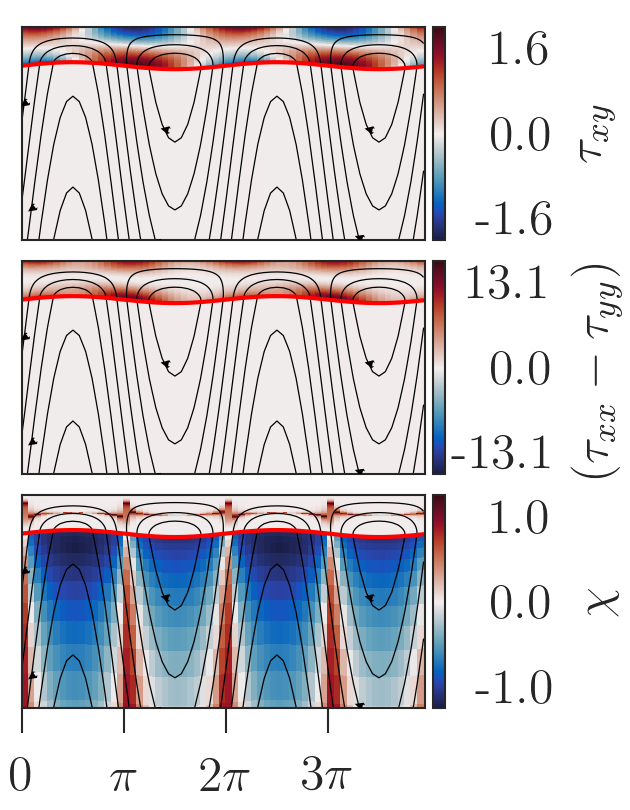}
    \end{minipage}
    \caption{\new{The polymeric stress, $\bs{\tau}$ surrounding a sheet with
            amplitude $bk = 0.01$ near a wall a distance $h_+k = 0.11$ above it
            in an Oldroyd-B fluid with $\beta = 0.5$ and $\De = 0.0$ (left),
            $\De = 0.5$ (middle) and $\De = 3.2$ (right).  The wall below the
            sheet is at $h_-k = 13.0$ which is far enough away to have no
            effect, however, the fluid domain is only shown until $ky = -0.5$.
            The swimming speed of the sheet in each situation is $U = U_N =
            0.0242c$ (left), $U = 0.905 U_N$ (middle), $U = 0.559 U_N$
            (right).} \label{fig:tauchi-small}}
\end{figure}

In line with the small-amplitude mechanism discussed above, we observe formation
of lines of strong extensional flow that generate large normal stresses
$\tau_{xx}-\tau_{yy}$, which are advected by the local flow. The main difference
between this case and the small-amplitude one is the fact that the normal
stresses generated in-between the wall vortices are now advected by the vortices
towards the sheet and also contribute to the shear stress $\tau_{xy}$ that
generates an additional average flow that, in turn, drags the sheet in the
direction of the wave. At small Deborah number, $\De=0.5$, the negative value
(blue) of the normal stresses $\tau_{xx} -\tau_{yy}$ is rotated into extra
positive (red) $\tau_{xy}$, which is responsible for the slow down of the sheet
relative to swimming in a Newtonian fluid.  This only happens in the vortices in
the troughs of the sheet, where the vortices are not restricted too much by the
presence of the wall. At larger Deborah number, $\De=3.2$, in addition to a
region with extra positive $\tau_{xy}$, there is a region with extra negative
$\tau_{xy}$ which pushes the sheet in the direction of its swimming, in
competition with the positive region.  The growth of this region of negative
polymeric shear stress is absent from the small amplitude solution \new{(shown
    in \cref{fig:tauchi-small})} and is responsible for the increased swimming
speed compared to the small amplitude prediction \cref{eq:small-speed}.

\begin{figure}
    \centering
    \begin{minipage}{0.45\textwidth}
    \includegraphics[width=2.9in]{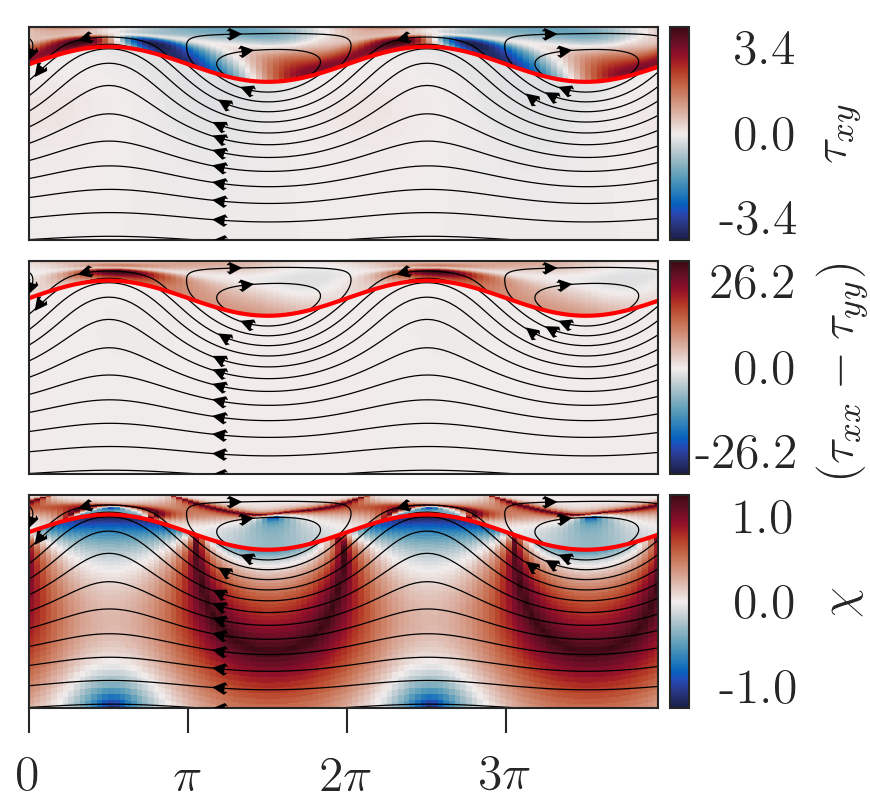}
    \end{minipage}
    \begin{minipage}{0.45\textwidth}
    \includegraphics[width=2.9in]{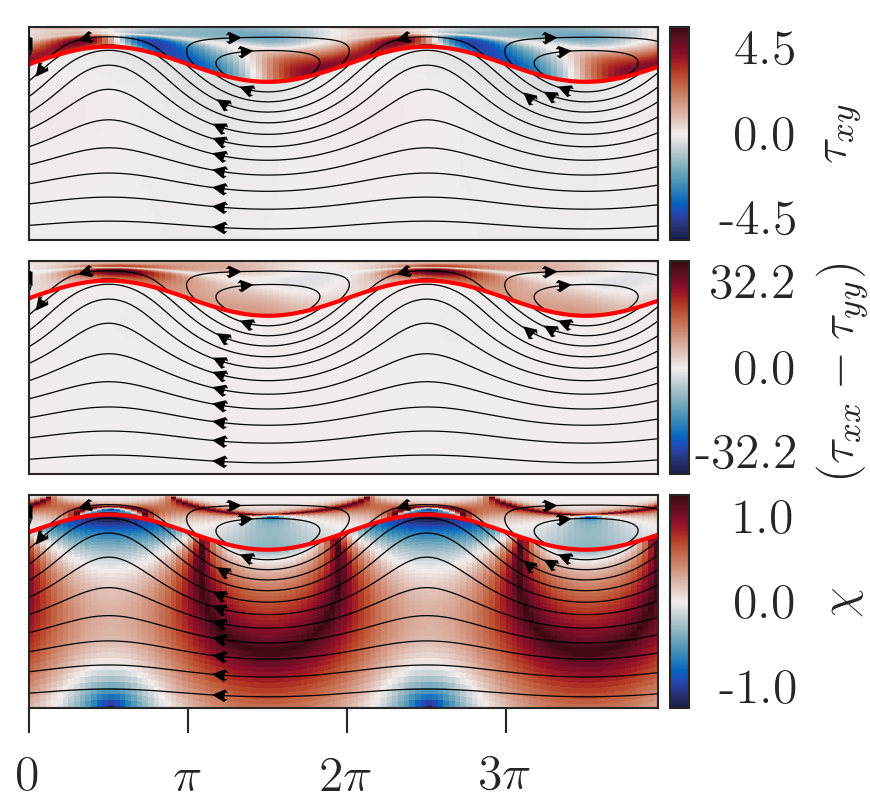}
    \end{minipage}
    \caption{Comparison between the stress distributions around a sheet with the
        amplitude $bk = 0.5$  at a distance $h_+k = 1.05$ from the upper wall
        with $\De=2.7$: $\beta = 0.5$ (left) and $\beta = 0.9$ (right). The wall
        below the sheet is at $h_-k= 13.0$ which is far enough away to have no
        effect, however, the fluid domain is only shown until $ky = -5.0$.
        The swimming speed in each case is $U=0.992U_N$ (left) and $U=1.020U_N$
        (right) with $U_N = 0.440c$.} \label{fig:tauchi2}
\end{figure}

Now we turn to the case of non-monotonic behaviour of the swimming speed with
the Deborah number.  In \cref{fig:tauchi2} we compare the stress distributions
for $\beta=0.5$ and $\beta=0.9$ for $h k = 1.05$ and $\De=2.7$. For $\beta=0.5$
these values approximately correspond to the local maximum of the swimming
speed, although its value is still smaller than the Newtonian one, while at
these parameters the case with $\beta=0.9$ exhibits swimming speeds larger than
$U_N$, see \cref{fig:large-b}. First, we note that now both the trough and crest
vortices are equally close to the wall, somewhat in contrast to
\cref{fig:tauchi}. This behaviour is also observed in the Newtonian case
$\De=0$, not shown. However, in  \cref{fig:tauchi2} there are no significant
differences between the stress distributions for $\beta=0.5$ and $\beta=0.9$
cases besides the numerical values of the stresses, and we conclude that whether
the swimming speed is larger or smaller than its Newtonian counterpart is
determined by a numerical competition between the wall and sheet stresses that
cannot be deduced from hand-waving arguments.

In conclusion, we have provided a mechanistic explanation for the
small-wave-amplitude swimming speed of a Taylor sheet derived by Elfring and
Lauga \cite{Elfring2015}, and explain why the effects of fluid's viscoelasticity
and the presence of a solid boundary decouple. We also developed a numerical
method with spectral accuracy that allows us to study finite-amplitude sheets of
various waveforms close to and away from solid walls. We observe that at finite
amplitudes the swimming speed is no longer a monotonic function of the Deborah
number, and can even become larger than the corresponding Newtonian value.
Interestingly, this effect seems to be the stronger, the more dilute the
viscoelastic solution is (large values of $\beta$), although there are
indications that at lower values of $\beta$ the swimming speed starts to
increase with $\De$ at sufficiently large values of the Deborah number. This
result suggests that even small amounts of polymer, either excreted or naturally
present in the solution, can aid propulsion next to solid boundaries, although
the speed increase reported in this work is minute. Our numerical data is not
sufficient to determine whether this increase would eventually lead to swimming
speeds larger than the Newtonian values for all values of $\beta$, at what
values of $\De$ this can be achieved, and how significant this speed up might
be. Further study is required to address these questions.

\section*{Acknowledgements}

Research outputs generated through the EPSRC grant EP/I004262/1 can be found at
http://dx.doi.org/xxxxxx.

\bibliographystyle{unsrt}
%\bibliography{references}
\bibliography{thesis}

\end{document}